\documentclass[aps,pra,superscriptaddress,twocolumn,showpacs]{revtex4-1}

\usepackage{amsmath, amsthm, amssymb}
\usepackage{graphicx}
\usepackage{color}
\usepackage{multirow}

\newcommand{\ket}[1]{{\left\vert{#1}\right\rangle}}

\usepackage{natbib}
\usepackage{hyperref}
\hypersetup{
        colorlinks=true,
}

\usepackage[scientific-notation=true,round-precision=3,round-mode=figures]{siunitx}

\newcommand{\eqnref}[1]{(\ref{#1})}
\newcommand{\ord}[1]{{\cal O}(#1)}
\newcommand{\braket}[2]{\langle #1 | #2 \rangle}
\newcommand{\be}{\begin{equation}}
\newcommand{\ee}{\end{equation}}

\begin{document}

\title{Gate count estimates for performing quantum chemistry\\ on small quantum computers}

\author{Dave Wecker}
\affiliation{Quantum Architectures and Computation Group, Microsoft Research, Redmond, WA 98052, USA}

\author{Bela Bauer}
\affiliation{Station Q, Microsoft Research, Santa Barbara, CA 93106-6105, USA}

\author{Bryan K. Clark}
\affiliation{Station Q, Microsoft Research, Santa Barbara, CA 93106-6105, USA}
\affiliation{Kavli Institute for Theoretical Physics, University of California, Santa Barbara, CA 93106, USA}

\author{Matthew B. Hastings}
\affiliation{Station Q, Microsoft Research, Santa Barbara, CA 93106-6105, USA}
\affiliation{Quantum Architectures and Computation Group, Microsoft Research, Redmond, WA 98052, USA}

\author{Matthias Troyer}
\affiliation{Theoretische Physik, ETH Zurich, 8093 Zurich, Switzerland}

\begin{abstract}
As quantum computing technology improves and quantum computers with a small but non-trivial
number of $N\ge100$ qubits appear feasible in the near future the question of possible applications
of small quantum computers gains importance.
One frequently mentioned application is Feynman's original proposal of simulating quantum systems, and in
particular the electronic structure of molecules and materials.  
In this paper, we analyze the computational requirements for one of the standard algorithms to
perform quantum chemistry on a quantum computer. We focus on the quantum resources required 
to find the ground state of a molecule twice as large as what current classical computers
can solve exactly. We find that while such a problem requires about a ten-fold increase in the number
of qubits over current technology, the required increase in the number of gates that can be coherently executed
is many orders of magnitude larger.
This suggests that for quantum computation to become useful for quantum chemistry problems, drastic algorithmic improvements will be needed.
\end{abstract}

\maketitle

\section{Introduction}

The excitement over quantum computation stems from the promise that quantum
computers will be able to solve problems for which classical computers don't have
enough resources.
The evidence for this comes from the discovery of 
quantum algorithms~\cite{Shor1994,kitaev1995,shor1997polynomial,hallgren2007polynomial,ortiz2001,aharonov2009polynomial,childs2003exponential} 
which, at least asymptotically, are exponentially faster than classical algorithms.  
This assures us that eventually, when sufficiently large quantum computers exist, they will fulfill
this promise. On the flip side, simple quantum algorithms have already been performed: for example,
the number 15 has been factored~\cite{Vandersypen2001} and the energy of an extremely simple molecule has been calculated~\cite{Lanyon2010,du2010}.
Although an important first step, these quantum calculations are still deep in the regime accessible 
to classical computers.  It is interesting, then, to explore what minimal resources are needed for quantum computers to solve problems 
that classical computers are unable to solve. In particular, 
this encourages us to explore problem instances which are just big enough to be outside the range of classical computers (say, for the next decade) 
and understand the quantum resources needed to solve these problems.  
We call these classically-intractable problems. 
In this work, we take up this task for the area of quantum chemistry.

Feynman's original proposal for a quantum computer~\cite{Feynman1982} was motivated by the exponential complexity of 
simulating many classes of  quantum systems on a classical computer. 
The wave function of $N$ 2-level systems, ({\em e.g.} $N$ spin-1/2 variables pointing up or down in a quantum magnet or $N$ 
spin-orbitals in a molecule each being occupied with either 0 or 1 electrons) 
lives in the Hilbert space $\mathbb{C}^{2^N}$ and thus needs an exponentially large number of $2^N$ classical variables to store. 
In contrast, on a quantum computer storing the same wave function requires only $N$ qubits. 
This reduces the memory requirement from exponential to linear and the runtime cost for many 
computations on the quantum system from exponential to polynomial.

The current state of the art in numerically exact classical algorithms, based on the diagonalization of the Hamiltonian matrix
using standard linear algebra methods either in the full Hilbert space or in a large Krylov subspace, can reach 
approximately $N=50$ spin orbitals \cite{nakano2011,lauchli2011ground,capponi2013numerical,gan2005calibrating}. 
Approximate methods for fermionic computation are starting to reach chemical accuracy on strongly correlated systems
for up to $N=70$ spin orbitals \cite{Kurashige2013}.
Hence, an interesting application of a quantum computer needs to reach at least $N=50$ spin orbitals to offer any advantages over
classical machines and realistically needs approximately $N=100$ spin orbitals to be significantly more useful than current
classical algorithms. To store a wave-function of this size requires full coherent control over at least 100 qubits. 
Experimentally, such systems seem feasible in the near-term future: Ion trap experiments have already demonstrated
coherence and entanglement between fourteen qubits~\cite{Monz2011} and many more ions have been trapped, but not yet entangled.
Using superconducting qubit technology, around 10 qubits can be controlled and a few hundred gates can be
executed coherently.

While there has been great progress towards non-trivial quantum computers with a small number of qubits, the 
development of quantum algorithms and exploration of applications for such devices has lagged behind. 
Factoring large integers using Shor's algorithm~\cite{Shor1994} is the canonical application for quantum computers, 
but it requires many thousands of qubits to factor a number that cannot be factored by classical
algorithms~\cite{beauregard2003,whitney2009fault,factorizationrecord}. 
The electronic structure problem for molecules, on the other hand, seems a more natural place where non-trivial
applications may exist for machines with a limited number of qubits and a significant amount of literature has
been devoted to this
topic~\cite{lidar1999,ortiz2001,AspuruGuzik2005,kassal2008,wang2008,whitfield2011,kassal2009,veis2010,
kassal2011,anmer2011,anmer2012,seeley2012,peruzzo2013,whitfield2013spin,yung2013,lamata2013,
babbush2013,CodyJones2012,kais2014};
for recent reviews, see Refs.~\cite{kassal2011,yung2012introduction}. 
Like Shor's algorithm, solving the electronic structure problem may also admit an exponential speedup 
but affords interesting possibilities with fewer necessary qubits.  
In addition, the technological benefits of quantum chemistry simulations are rich: 
For example, finding better catalysts to be used in many industrial-scale chemical processes -- 
even including very basic processes, such as nitrogen fixation -- has challenged researchers for decades~\cite{Podewitz2011}. 
In these problems, approximate approaches such as density-functional theory do not yield
sufficient accuracy for the correlation energies, while more accurate methods, such as the
density matrix renormalization group, have so far not been able to simulate sufficiently large
systems in a reasonable time-frame.

While there are many important quantities that characterize molecular systems, in this paper, we focus 
particularly on the measurement of the 
electronic ground state energy. Computing these energies (and their respective derivatives) is a 
basic starting point for computing other observables. We implement one of the standard approaches
to performing such calculations on a quantum computer~\cite{AspuruGuzik2005,whitfield2011}, for brevity henceforth
referred to as {\it quantum full configuration interaction} (QFCI), at the level of the individual circuit
elements and compare the results obtained for a water molecule 
(10 electrons, 14 spin-orbitals in an STO-3G basis) to those obtained by the equivalent standard full
configuration interaction (FCI) calculation, validating the approach. We note that water in this basis
is a standard example that was already considered in Ref.~\onlinecite{AspuruGuzik2005}. We are then able to 
measure the costs involved in the quantum computation, including the number of qubits, number
of circuit elements, and the circuit depth needed to  perform this computation.

By systematically analyzing the effects of time-step errors and gate count, we show the scaling of the algorithm with the
number $N$ of spin orbitals to be $\ord{N^9}$ if all gates are executed in a serial fashion, and $\ord{N^8}$ if we allow
for parallel execution of gates. Using the example of a water molecule, we also determine
the prefactors involved and thereby set a baseline of requirements for a quantum computer to
perform calculations on larger molecules.

\section{The Coulomb Hamiltonian in Quantum Chemistry}

Using a Born-Oppenheimer approximation to fix the positions of the nuclei in space,
the electronic structure problem for a molecule reduces to finding the low-lying spectrum
of the electronic degrees of freedom. For a full configuration interaction approach, a
basis of single particle orbitals, such as the STO-3G basis used here, must be chosen. The choice
of basis here dictates the number of orbitals that need to be considered in the FCI calculation. One
then obtains a basis of molecular orbitals by performing a Hartree-Fock calculation and rewrites
the Hamiltonian in a second-quantized form in terms of these orbitals, where it takes the form
\begin{equation} \label{eqn:H}
H = \sum_{pq} t_{pq} c_p^\dagger c_q + \frac{1}{2}\sum_{pqrs} V_{pqrs} c_p^\dagger c_q^\dagger c_r c_s.
\end{equation}
Here, $c_p$ and $c_p^\dag$ denote the annihilation and creation operators for an electron in a set of $p=1,\ldots,N$ spin orbitals.
For this paper we want to focus on calculating the ground state energy $E_0$ and generating the ground
state wave function $\ket{\psi_0}$ of this Hamiltonian.

The energy scales for the molecular problems are set by the core energy of the atom, giving energies of approximately
100-1000 Hartree ($E_h$) for small molecules. In order to 
perform useful quantum chemistry, a chemical accuracy of approximately 1~milli-Hartree ($mE_h$) is important.  
This means we have to resolve the energy scales to one part in a million.  

Currently, quantum chemists use the exact full configuration interaction (FCI) method, and a variety
of approximate methods to tackle this problem. 
Some of the approximate methods are quasi-exact in the sense that they can be systematically improved by increasing some accuracy parameter or by accumulating statistics longer; 
these include the density-matrix renormalization group~\cite{white1992,white1992-1,marti-book,chan2011}, other tensor network states~\cite{marti2010}, 
quantum Monte Carlo methods such as FCIQMC~\cite{booth2009,cleland2010,booth2013}, and coupled cluster (CC)~\cite{cizek1966}.   While some of these approaches 
have the virtue of even scaling polynomially in the desired accuracy, none of these approaches scale polynomially in molecule size for the generic 
molecular system.   Other, more widely used approximate methods cannot be systematically refined, such as density functional theory (DFT)~\cite{hohenberg1964,kohn1965}, 
but permit the study of much larger molecules of up to thousands of atoms.  

\section{Quantum Full Configuration Interaction algorithm}
We now outline the algorithm we use to determine the ground state energy of a small molecule. This
algorithm has been previously described in several papers~\cite{whitfield2011}.
The first step is to prepare the qubits into a state $\ket{\psi}$ which is a good approximation to the
ground state $\ket{\psi_0}$, e.g. it has sufficiently high overlap $\braket{\psi}{\psi_0}$. 
For a sufficiently small molecule -- like the water molecule used here -- this can just be the Hartree-Fock 
solution $\ket{\psi_\text{HF}}$. By choosing as basis functions the single-particle wave functions obtained in a Hartree-Fock 
calculation we  can write $\ket{\psi_\text{HF}} = \prod_{i=1}^{N_e} c^\dagger_i \ket{0}$, where $\ket{0}$ 
is the vacuum, $N_e$ the total number of electrons, and $c^\dagger_i$ creates an electron in the $i$'th single-particle
state.

For larger molecules we expect the overlap $\langle \psi_\text{HF} \ket{\psi_0}$ to decrease significantly. In that case a better approximation
to $\ket{\psi_0}$ can be obtained by adiabatic evolution of the wave function from $\ket{\psi_\text{HF}}$
towards the true ground state $\ket{\psi_0}$. This can be achieved by evolving the wave function under the action
of a Hamiltonian which slowly evolves from the Hartree-Fock Hamiltonian $H_{\rm HF}$ to the 
full Coulomb Hamiltionian (\ref{eqn:H})~\cite{AspuruGuzik2005}. Here, the initial Hamiltonian must meet the requirement
that it has the Hartree-Fock state as its unique ground state.

In the second part of the QFCI algorithm, the energy of the state $\ket{\psi}$ obtained through the above
preparation procedure is measured using the {\it quantum phase estimation} (QPE) algorithm~\cite{kitaev1995,kitaev2002book}.
This also collapses the state $\ket{\psi}$ (with probability proportional to $|\braket{\psi}{\psi_0}|$) to the ground state wave-function.
We note that there are proposals to improve the quantum phase estimation algorithm~\cite{Svore2013} as
well as proposals to avoid it altogether in the quantum chemistry context~\cite{peruzzo2013}.

At the highest level, quantum phase estimation takes a state 
$\ket{\psi}=\sum_i c_i \ket{\phi_i}\ket{0}$, where $\ket{0}$ is the
initial state of a number of auxiliary qubits, and converts it into the state
$\sum_i c_i \ket{\phi_i}\ket{E_i}$, where $\ket{E_i}$ denotes that a binary
representation of the energy has been encoded into the auxiliary qubits.
It is very important to note that at the core of the algorithm lies the time evolution
of a quantum state, i.e. performing $\ket{\psi(t)}= \exp( -i H t) \ket{\psi(0)}$. The
time $T$ required to resolve an absolute error in the energy $\epsilon$ is $\pi/\epsilon$.

On a general-purpose quantum computer, time evolution must be implemented through
a circuit composed of one and two-qubit gates. In certain algorithms, most notably Shor's
algorithm, the time evolution $\exp( -i H t)$ can be implemented efficiently by exploiting
special properties of the evolution operator, such that the computation time necessary
for the whole time evolution scales as $\ord{\log t}$.
In general, however, a different approach must be taken, the most common one being
a Trotter decomposition; other approaches exist, however~\cite{anmer2012,peruzzo2013,babbush2013}.
In a Trotter decomposition~\cite{trotter1959,suzuki1976}, the full time $t$ is divided into discrete time intervals
$\Delta_t = t/M$. This leads to a scaling that is at least linear in $t$, and it
incurs a discretization error that is polynomial in $\Delta_t$, with the exponent depending
on the type of Trotter decomposition used.

For a Hamiltonian $H$ which is given as a sum over individual terms $h_k$, the first order 
Trotter decomposition with $M$ Trotter steps reads
\begin{eqnarray} \label{eqn:trotter}
\exp( -i t\sum_k h_k) &=& U(\Delta_t)^{M} + \ord{ \Delta_t } \\
{\rm with} \qquad U(\Delta_t) &=&  \prod_k \exp (-i \Delta_t h_k ).
\end{eqnarray}
A representation of the Hamiltonian must be chosen where each term
$U_k(\Delta_t) = \exp (-i \Delta_t h_k)$ can be broken down into a sequence of standard gates;
this is the case for example if each $h_k$ is a product of Pauli matrices.
At the cost of a factor of two in the number of circuit elements, a second order Trotter decomposition can be used, which
improves the error to $\ord{ \Delta_t^2}$.  In principle, even higher order Trotter breakups can further attenuate the error. 
Notice that the time step error coming from using phase estimation with an approximate time evolution operator $U$ is exactly equivalent to 
an error-free approach with the effective Hamiltonian  $H_\textrm{eff}=\ln U(\Delta_t)/(-i\Delta_t).$
The circuit diagrams for all terms that occur in the quantum chemistry
Hamiltonian~\eqnref{eqn:H} have been previously obtained~\cite{whitfield2011}, but to make this presentation
self-contained are shown in Appendix~\ref{sec:circuits}.  

In the rest of this paper, we focus solely on the second part of the QFCI algorithm,
i.e. measurement of the energy through quantum phase estimation. To compute the
computational effort involved in this algorithm, the three factors
that need to be taken into account are the number of gates per Trotter step $N_g$, 
the total time $T$ that the time evolution must evolve in phase estimation,
and the total number of Trotter steps $1/\Delta_t$ needed for evolving for fixed time at fixed error.
The total complexity is then $N_g T /\Delta_t$. 
The total time $T$ is set by the absolute accuracy required. Using an absolute accuracy of 1~milli-Hartree,
we get that $T\approx 6000~E_h^{-1}$ is required. 

We  proceed by computing the $\Delta_t$ and $N_g$ required for water and then extrapolate to 
classically-intractable systems.

\section{Results -- Water}

We implement the quantum phase estimation algorithm for a water molecule in a minimal STO-3G basis of ten electrons in 
fourteen spin-orbitals. We first perform a Hartree-Fock calculation using the PyQuante package~\cite{PyQuante} and use 
the thus obtained orthonormalized single particle wave functions as the basis used in the quantum algorithm.
We validate our
implementation of the quantum algorithm by comparing to full-configuration-interaction (FCI) calculations for the same problem.

Our simulations are performed using the LIQ$Ui\ket{}$ quantum simulation platform~\cite{LiquidOverview},
which is an advanced software package developed by Microsoft 
Research to allow efficient simulation of large quantum circuits (more than 1 million gates) with moderate numbers of 
qubits (typically 30 qubits in 32GB of memory for Hamiltonian simulations).
The system is implemented as an extension to the F\# functional programming language~\cite{fsharp} and compiles high-level circuit 
descriptions into targeted simulators (Universal, Stabilizer and Hamiltonian) in one of several environments (Client, Service or Cloud). 
The architecture is modular and includes packages for optimization, noise modeling, physical gate replacements~\cite{dawson2006},
export and automatic circuit drawing \footnote{all circuits in this paper were auto-generated by LIQ$Ui\ket{}$}.
For this work, a module was added to convert terms into individual circuits, analyze and optimize rotations and then compile 
re-written circuits into unitary matrices for analysis and simulation.

\begin{figure}[t]
  \includegraphics[width=3in]{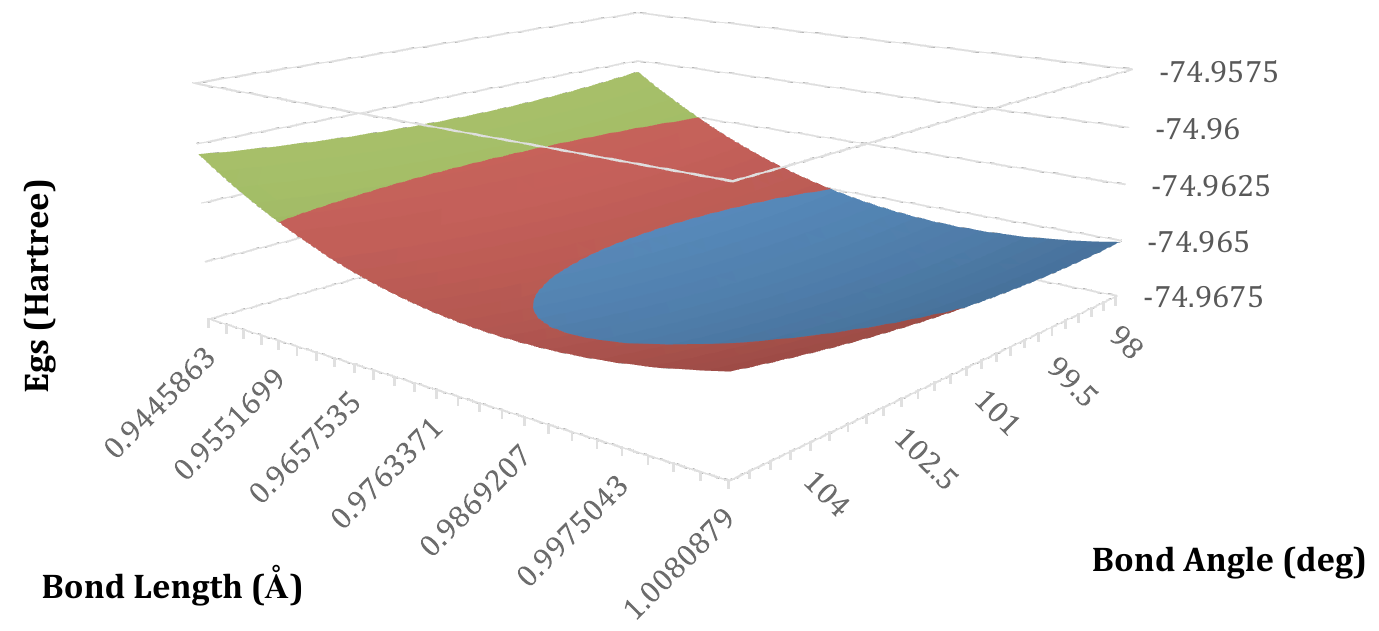}
  \caption{(Color online) This figure shows the energy of the water molecule as a function of bond angle and bond length for an STO-3G basis obtained from
  a restricted HF calculation. \label{fig:h2oenergy} }
\end{figure}

\begin{figure}[t]
  \includegraphics[width=\columnwidth]{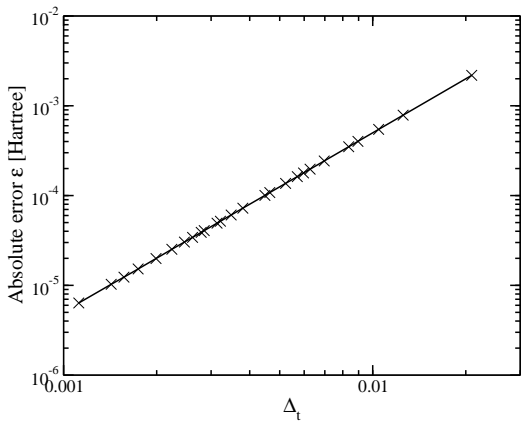}
  \caption{Discretization error due to the Trotter decomposition for a water molecule at bond length $a=0.957213~\mathrm{\AA}$ and bond angle $\theta=104.5225^{\circ}$. \label{fig:trotter} }
\end{figure}

In Figure~\ref{fig:h2oenergy}, we show the energy of a water molecule as a function of bond length and bond angle as obtained from our 
simulated QFCI calculation. Figure~\ref{fig:trotter} shows the dependence of the accuracy compared 
to an exact solution of the same problem on the Trotter time step.
We find that a Trotter time step of $\Delta_t = 0.01~E_h^{-1}$ is required to achieve chemical accuracy. 
We find that, for most orderings of the
terms in the Hamiltonian, the error behaves as $\ord{\Delta_t^2}$ even for the first-order Trotter decomposition of
Eqn.~\eqnref{eqn:trotter}. We attribute this to a cancellation in the errors, which seems to be fairly generic.
Also, in this regime it does not seem significantly advantageous to go to a higher-order Trotter decomposition. Considering
just the scaling of the error, choosing a fourth-order decomposition would allow us to take
$\Delta_t = 0.1~E_h^{-1}$ and would thus lead to a ten-fold decrease in the number of Trotter steps,
but at the same time would increase the number of gates for a single Trotter
step by a larger factor~\cite{efficient2007}. This trade-off may change in different parameter regimes,
for example if a much smaller $\Delta_t$ is required.
The convergence of our results to the FCI values confirms the correctness of our implementation of simulated QFCI
and goes a step beyond the simulations performed previously~\cite{AspuruGuzik2005,veis2010,whitfield2011,seeley2012}.

We now turn to estimating the resources for this simulation on a quantum computer. To this end, we quantify
both the total number of circuit elements and what we refer to as the {\it parallel depth} of a circuit, which is
the minimum required depth of the circuit when exploring possible parallelism between parts of the circuit
that operate on disjoint sets of qubits and can hence be executed simultaneously.
We also separately count the rotations required as these are the most costly operations for many physical
realizations of a quantum computer.
We can group the terms in the Hamiltonian into different categories, each requiring a different number of
elementary gates.
For a fermionic problem, Jordan-Wigner strings~\cite{Jordan1928} are generally used to enforce fermionic
signs. These will increase the number of gates necessary to apply an off-diagonal term by a factor of
$N$. Recently, methods have been developed that can improve this~\cite{bravyi2002,CodyJones2012,seeley2012}; for example, Ref.~\cite{CodyJones2012}
describes a method where Jordan-Wigner strings can be applied in constant time at the cost of $N$ additional
teleportations, which however can be carried out in parallel.
It is at this point unclear whether the teleportation can be executed at a similar clock rate as gate operations;
nevertheless, we include this possible improvement in our gate counts for parallel operations.

Using the actual gate counts for each term, shown in Table~\ref{fig:circuitdepth} in Appendix~\ref{sec:circuits}, we 
calculate the circuit depth for one Trotter time step and show it in Table~\ref{table:gatecounts}. 
We find that for our water simulation, the gate count is $20494$ for sequential operations and
$6438$ for parallel operations.
Given the required time step $\Delta t= 0.01~E_h^{-1}$ and total time $T=6 \cdot 10^3~E_h^{-1}$ we must evolve for the QPE to achieve chemical accuracy, which we have established through our simulations above, we need approximately $6 \cdot 10^{5}$ Trotter steps and $10^{10}$ serial gates. In this case, working in parallel leads to a speedup of a factor of 3.

\section{Scaling to larger molecules}

Having set a baseline using our simulations of the water molecule, we need to examine the
scaling of two key quantities to be able to extrapolate our results to larger molecules:
(i) the number of gates $N_g$ needed for a single Trotter step,
(ii) the value of the Trotter time step $\Delta t$.

\begin{figure}
  \includegraphics[width=3in]{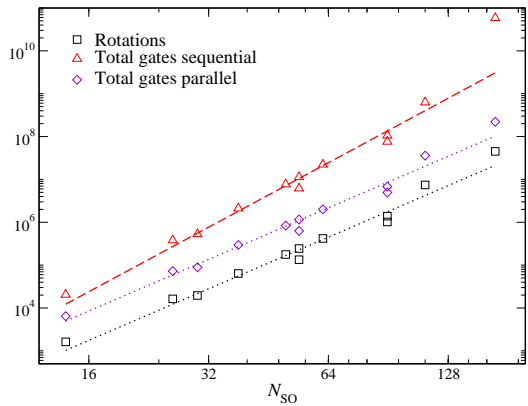}
  \caption{(Color online) (i) Number of rotations, and number of gates in (ii) parallel and (iii) sequential mode.
  Dashed and dotted lines indicate the expected scaling, which is $(N_{SO})^4$ (dotted lines) for the
  number of rotations and the parallel gate count, and $(N_{SO})^5$ (dashed line) for the sequential gate count.
  This data is also listed in Table~\ref{table:gatecounts}.
  \label{fig:gatecountscaling} }
\end{figure}

We anticipate the number of gates to be proportional to the number of terms in the Hamiltonian,
$\ord{N^4}$, multiplied by the number of gates for each term, which due to the Jordan-Wigner
transformation is roughly $N$, thereby giving a scaling of $\ord{N^5}$. Note that this assumes that
all gates have roughly the same cost; in reality, this may be drastically different. In particular, the
number of rotations, which are likely the most expensive gate, does not depend on the Jordan-Wigner
strings and will therefore be at most $\ord{N^4}$. Also, as mentioned before, there have been proposal
to reduce the cost of Jordan-Wigner strings from $\ord{N}$ to $\ord{1}$~\cite{bravyi2002,CodyJones2012,seeley2012}.
In Fig.~\ref{fig:gatecountscaling} and Table~\ref{table:gatecounts}, we show the scaling of the number
of rotations as well as the number of gates in parallel and sequential operation. We find that the empirical scaling matches 
our expectations quite accurately.

\begin{figure}
  \includegraphics[width=\columnwidth]{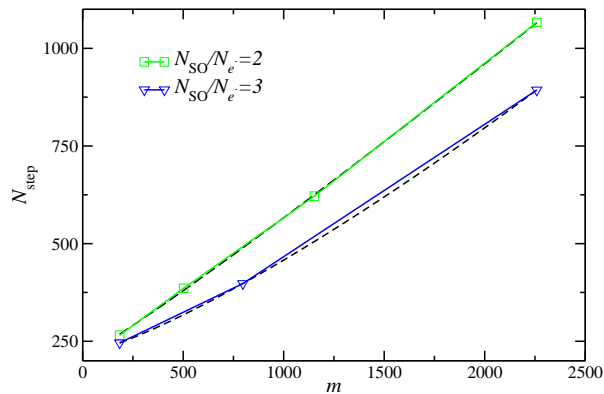}
  \caption{(Color online) Scaling of the number of necessary Trotter steps, $1/\Delta t$, with the number of terms in the Hamiltonian $m$. This is extracted
  from a series of artificial, but statistically appropriate Hamiltonians with 12, 16, 18, 20 and 24 spin-orbitals. The two curves correspond
  to different values of the inverse filling $r=N_\text{SO}/N_{e^-}$. The dashed lines indicate fits to $N_\text{step} \sim m^\alpha$, with exponents
  $\alpha(r=3)=1.27$ and $\alpha(r=2)=1.08$. \label{fig:dtscaling} }
\end{figure}

Secondly, we need to establish how the Trotter step necessary to attain a given accuracy scales with the system size.
Theoretical calculations bound the number of Trotter steps required for a fixed time $T$ by $m^{1+1/2k}$~\cite{efficient2007},
where $m$ is the number of separate terms the Hamiltonian  $H=\sum_{i=1}^m H_i$ is split into, and $2k$ is the order of the
Trotter decomposition; see Appendix~\ref{sct:app-commutator} for the error in ground state energy due to Trotter error.
This theorem of Ref.~\onlinecite{efficient2007} is stated in terms of the operator norm of $H$ (written as $\Vert H \Vert$) which would depend upon $N$; however it seems from the derivation that the bound in fact depends upon $\Vert H_i \Vert$ which for these synthetic molecules is chosen independent of $N$.
In our case, we effectively use a second-order decomposition and have $\ord{N^4}$ terms,
such that the number of Trotter steps required is bounded by $\ord{N^6}$. Since this is only a bound and since we are
not necessarily in the asymptotic regime, it is important to test the scaling  empirically.
We do this by producing a series of artificial molecules whose Hamiltonian terms have the same
statistical properties as real molecules and measuring the scaling as a function of the number of terms in these molecules.
Our result is shown in Fig.~\ref{fig:dtscaling}; details are discussed in Appendix~\ref{sct:app-scaling}.
By performing a fit to the data (shown as dashed line in the figure), we obtained exponents of 1.27 or 1.08,
depending on the ratio of the number of electrons to the number of spin orbitals. Hence, we find
that the scaling is closer to $N_\text{step} = 1/\Delta_t \sim m$ instead of $m^{3/2}$,
and we have $1/\Delta_t \sim N^4$.

One possible reason for this improvement is that each term in the Hamiltonian has a nonvanishing commutator with at most $\ord{N^3}$ terms, which allows us to give an improved bound on the number of Trotter steps required.  See Appendix~\ref{sct:app-commutator} for this bound.

\begin{table}
  \begin{tabular}{|c|c|c|c|c|} \hline
  &$N_t$ &$\frac{\text{Gates}}{\text{term}}$ &$1/\Delta_t$ &Total (Parallel) \\ [1.1ex]\hline \hline
  Upper bound &$N^4$ &$N$ &$(N_\text{t})^{3/2} = N^6$&$N^{11}$ ($N^{10}$) \\ \hline
  Empirical scaling &$N^{3.8}$ &$N$ &$N^4$ &$N^{9}$ ($N^{8}$)\\ \hline
  \end{tabular}
  \caption{Summary of the different contributions to the overall scaling. Here,  $N_t$ is the number of terms; gates/term denotes the number of gates required to execute a term using a basic Jordan-Wigner transformation, and assuming all gates take an equal amount of time; $\Delta_t$ is the Trotter time step. \label{table:scalings} }
\end{table}

We have summarized all these contributions in Table~\ref{table:scalings}. We find that the total scaling goes
approximately as $\ord{N^{9}}$ for sequential operation.
Note that this is better then expected from the most general bounds, which would give a scaling of $\ord{N^{11}}$.
As an example, we take Fe$_2$S$_2$ in the STO-3G basis; this molecule has been considered a benchmark
example and has enormous importance to biochemistry.
Given that the basis of 112 spin orbitals is about 8 times larger than the basis for H$_2$O,
we see that we pay a factor of $8^5$ from the number of gates per Trotter step, 
and a factor of $8^4$ from the smaller Trotter step required, leading to a total
increase of the runtime by a factor of approximately $10^8$; we thus have to
execute a total number of $10^{18}$ gates. For Fe$_2$S$_2$, working in parallel
(including the execution of Jordan-Wigner strings in constant time)
gains us a factor of 20 leaving us with a parallel gate depth of $10^{17}$.
An additional factor may be gained by choosing a higher-order Trotter decomposition,
which may be optimal for larger molecules~\cite{efficient2007}.

\subsection{Quantum hardware requirements for classically-intractable molecules}

From the perspective of the number of qubits, the hardware required to simulate a classically-intractable
molecule is within reach: current technology can operate on about a dozen qubits, which is about
an order of magnitude less than the 100 qubits that are required for the Fe$_2$S$_2$ molecule that
we have used as an example above. However, achieving the necessary gate count seems much
more challenging: current technology allows control for a few hundred gate operations, while we have obtained
a gate count of $10^{18}$ as an upper bound to the number of gates required to simulate Fe$_2$S$_2$.
An improvement of more than 15 orders of magnitude over this bound is therefore necessary!
Indeed, even the calculation for the ground
state of water, which takes mere seconds on a classical computer and requires only few
more than a dozen qubits, is many orders of magnitude too large in terms of the gate-count.
Even if we assume that Moore's law applies to quantum computers, i.e. the number of gates
that can be executed doubles every 18 months, it will take 75 years to be able
to simulate Fe$_2$S$_2$ using the QFCI algorithm as discussed here.
Even then, assuming a clock
speed of 1 GHz (i.e. 1 ns per gate) for gate operations, the calculation for
for Fe$_2$S$_2$ will require 1.5 years to complete!
We note that these estimates require the quantum state to be coherent for the entire calculation as the no-cloning
theorem forbids checkpointing intermediate results. 

Moreover, the numbers we have cited so far are given in terms of {\it logical} qubits and gate operation,
i.e. ideal qubits and gates that are perfectly coherent and operate with perfect fidelity. In reality,
qubits have a finite coherence time and gates can only be executed with less than perfect fidelity. Therefore,
in most physical implementations, a logical qubit will have to be represented through a number of
{\it physical}, i.e. hardware qubits, and a logical gate operation is obtained via a series of physical
gate operations. This allows the use of error correction. In conventional
error correction schemes, depending on the fidelity with which the physical gate operations
can be executed, between 100 and $10^5$ physical gate operations are required for a
single logical gate operation; however, it is possible to trade off the number of gates with the number
of extra qubits required~\cite{gottesman2009introduction,devitt2013,gottesman2013}.
The alternative route of topological quantum computing~\cite{nayak2008}, where the physical realization of
the qubit itself is robust against errors, would require braiding, i.e. adiabatically moving
excitations of the underlying topological phase, to operate at least an order of magnitude faster than the required
logical gate clock rate.
We have also ignored the computation time involved in preparing a state with overlap better then the
Hartree-Fock state and the possibility that the whole algorithm might
need to be repeated many times to accumulate statistics. 

In current technology, typical gate times are 10$\mu$s for ion traps and 100~ns (10 megahertz) for superconducting 
qubits. We note also that the speed at which a quantum computer can run is bounded by the speed at which 
we can do classical control and it is hard to see going beyond tens of gigahertz  
in the foreseeable future.

We thus conclude that the simulation of a molecule like Fe$_2$S$_2$ on a quantum computer using
the QFCI algorithm described in this paper will not be possible by hardware improvements alone;
instead,
algorithmic improvements of several orders of magnitude will be necessary.

\section{Conclusions}

In this work, we have answered the question of whether a small quantum computer with on the order of
100 qubits will be able to address challenging problems in quantum chemistry that are beyond the reach
of classical algorithms. From a purely conceptual point of view, it seems very likely that quantum computers
can achieve this: such a quantum computer could trivially represent the wave function of a molecule with up
to 100 spin orbitals, and an algorithm (referred to here as QFCI algorithm) is known that in principle should
allow the calculation of the ground state energy in polynomial time.
Exploring the details of this algorithm, however, we have been able to demonstrate that its polynomial scaling
is very large, and that the prefactors work out such that, under reasonable assumptions
about improvements in quantum computing, the classically-intractable regime remains
intractable also for a quantum computer.

We feel that these estimates draw a line in the sand setting an important barrier 
that must be overcome for the dream of useful quantum computation for the electronic
structure problem in molecules to become a reality.
Given even very optimistic assumptions, it seems unlikely that the
challenges presented in this paper can be overcome by hardware improvements.
Instead, we feel that this emphasizes the importance of {\it quantum software engineering}, i.e. the necessity
of algorithmic advances with a strong focus on practically applicable algorithms. In the realm of
classical algorithms, challenging problems are becoming tractable not only due 
to the fast increase of computational power, but even more so due to advances in algorithms.
A prominent example of such an advance are Monte Carlo algorithms with non-local updates, which
have lead to performance advances over the original Metropolis algorithm of many orders of magnitude.
We believe that similar advances in quantum algorithms will ultimately bring to fruition Feynman's
intuition~\cite{Feynman1982} that quantum computers should be better than classical computers at simulating
the properties of quantum systems.

\subsection{The Path Forward}

Having focused on the standard QFCI algorithm for solving the QC problem in a given basis
of molecular orbitals,
we can ask whether there exists alternate algorithms which might perform better in this regime.
For some alternative approaches, see also Refs.~\onlinecite{peruzzo2013,babbush2013}.
There has been significant work in the literature discussing the more general sparse Hamiltonian problem
where algorithms are designed to time-evolve (otherwise structure-less) sparse Hamiltonians.
The two current algorithms which scale best (in an incomparable way) are those of
Refs.~\onlinecite{berry2013exponential,berry2012black}.
In the former work, an algorithm is given that scales with the number of non-zero elements per column $d$ and
the total time $t$ as $\ord{d^2t\log^3(dt)}$, with a complexity polynomial in the logarithm of the inverse error.
In our quantum chemistry Hamiltonian, $d=N^2n_e^2$ where $n_e$ is the number of electrons.
Assuming $n_e$ scale as $N$, we get $\ord{N^8\log^3 N}$.
This is asymptotically better than the naive Trotter decomposition used in our QFCI algorithm,
with a crossover due to the $\log^3 N$ contribution which is naively at $N \approx 100$.
However, the constants may be very different and the bound may not be tight, such that
this algorithm may or may not be better than our algorithm discussed in this paper.
In the latter work of Ref.~\cite{berry2012black}, the authors use a quantum walk approach to simulate the Hamiltonian
and obtain a scaling of $\ord{d^{2/3} ((\log \log d) t \Vert H \Vert)^{4/3}}$ with bounded error,
where $d$ is the number of non-vanishing elements per row of the matrix. In our problem, $d=N^4$
and therefore we obtain the scaling $\ord{N^{8/3}((\log\log N^4) \Vert H \Vert)^{4/3}}$.
Here, $\Vert H \Vert$ is the operator norm of the Hamiltonian, which scales at least linearly
with the number of electrons (giving at least an $N^4$ total scaling), but may scale faster if correlation energies dominate. Alternately, Ref.~\cite{berry2012black} gives a quantum walk algorithm requiring time $\ord{d \Lambda_{max} t}$, where $\Lambda_{max}$ is the largest matrix element of the Hamiltonian.  This gives at least $N^4$ scaling also if $\Lambda_{max}$ is independent of $N$.
Both of these algorithms
require oracle access to matrix elements, which may incur an additional factor of up to $N^4$ in gate count to encode the coefficients of the Hamiltonians; some of these gates can be executed in parallel, at the cost of additional qubits. It remains to be explored whether this algorithm
can be used favorably in some parameter regime for electronic structure calculations.
Ongoing work is exploring how these tradeoffs work out in the relevant regime of classically-intractable molecules.

Beyond these, or other new algorithms, we can examine whether 
potential incremental improvements 
might chip away at the factors described in this work
gaining enough factors of $N$ to make a variant of QFCI tractable.

One problem with the current approach is the need to take \emph{many} time-steps.
To decrease this number, a higher-order Trotter decomposition is often suggested. Naively, this is 
problematic as it comes with significant overhead, but recent work~\cite{childs2012product,childs2012hamiltonian} 
has explored approaches to mitigate this. From a theoretical 
perspective, even an `arbitrarily high-order' Trotter decomposition
changes the scaling with the number of terms $m$ in the decomposition of the Hamiltonian from $m^{1+1/2k}$
to $m$. As we are already empirically seeing (and accounting in our estimate) a scaling of $m$, it is unclear if a  
higher order Trotter decompositions will garner significant gains; this said, it is
possible that these two effects would combine potentially saving a factor 
of order $N^2$. This assumes improvements make these `higher order' decompositions 
as cheap as a second order decomposition. In a similar vein, it may be possible to extrapolate the time
step error or have it cancel out in observables of interest giving us the ability to work at much
larger time steps.  Finally, we propose an adaptive trotter scheme in Appendix \ref{sec:coalescing} 
which might require significantly less gates per time step in certain regimes. 

Another possibility is to decrease the number of terms $m$ in the decomposition of the Hamiltonian.
For an example of where this has been done, see Ref.~\cite{childs2011simulating}.
Alternately, a different basis may be used.
In a local basis the total number of terms often scale as $N^2$ instead of $N^4$. This
would change the scaling of the complete algorithm from $\ord{N^{9}}$ to $\ord{N^5}$.  
In the extreme limit of a real-space basis, there may be significant additional gains coming from the fact
that the Hamiltonian can be decomposed into only two pieces \cite{kassal2008}. Working 
in real space brings its own independent set of problems, though, and we are currently looking into
whether this is a superior approach in the classically-intractable regime.

\begin{table*}
  \begin{tabular}{||c|c|c|c||c|c||c|c||}
  \hline & & & & \multicolumn{2}{c||}{Sequential} & \multicolumn{2}{c||}{Parallel}  \\
  \hline Molecule & Basis &  Spin orbitals & Basis size & Rotations & Total & Rotations & Total \\
  \hline
  \hline
  \multirow{6}{*}{H$_2$O}
 & STO-3G & 14 & 441 & \num{1615} & \num{20494} & \num{1615} & \num{6438}  \\
 & 3-21G & 26 & $1.66\times10^6$ & \num{16253} & \num{381208} & \num{16253} & \num{72536}  \\\
 & DZVP & 38& $1.35\times 10^8$ & \num{63863} & \num{2124678} & \num{63863} & \num{295430}  \\
& 6-31G** & 50&  $2.8\times 10^9$ & \num{177205} & \num{7694840} & \num{177205} & \num{837480}  \\
 & 6-31++G** & 54& \num{} $6.5\times 10^9$ & \num{244123} & \num{11404322} & \num{244123} & \num{1159082}  \\
 & 6-311G** & 62& \num{} $2.9\times 10^{10}$ &\num{419219} & \num{22330186} & \num{419219} & \num{2003978}  \\
\hline
  \multirow{4}{*}{CO$_2$}
  & STO-3G & 30 & $1.86\times 10^6$ &\num{19639} & \num{531926} & \num{19639} & \num{89062}  \\
  & 3-21G & 54 & $1.70\times10^{14}$ & \num{134231} & \num{6187878} & \num{134231} & \num{630006}  \\\
  & DZVP & 90& $1.03\times 10^{20}$ & \num{1394669} & \num{106047976} & \num{1394669} & \num{6791752}  \\
& 6-311G** & 90& \num{} $1.03\times 10^{20}$ & \num{1023013} & \num{76179240} & \num{1023013} & \num{4950192}  \\
\hline
 \multirow{2}{*}{Fe$_2$S$_2$}
 & STO-3G & 112 & $3.4\times 10^{25}$ & \num{7441260} & \num{630767773} & \num{7441260} & \num{35865821} \\
 & 3-21G & 168 &$2.8\times 10^{48}$ & \num{45074552} & $5.84 \times 10^{10}$  & \num{45074552} & \num{220771169}  \\
\hline
  \end{tabular}
  \caption{Gate count for one Trotter step using a sequential or a parallel circuit. The basis sizes are extracted from
  the PyQuante package~\cite{PyQuante}; for details on the basis sets, see also Refs.~\cite{bse,feller1996role,schuchardt2007basis}.
  \label{table:gatecounts} }
\end{table*}

\acknowledgements

We acknowledge useful discussions with M. Reiher, K. Svore, M. Freedman, A. Childs and the participants of the workshop {\it Quantum Computing for Quantum Chemistry} at Microsoft Research, Redmond, held in November 2012. We acknowledge hospitality of the Aspen Center for Physics, supported by NSF grant 1066293.
This research was supported in part by the National Science Foundation under Grant No. NSF PHY11-25915.

After completion of this work, improvements on the methods and bounds discussed here have been proposed in Refs.~\onlinecite{toloui2013,hastings2014,poulin2014}.

\appendix

\section{Scaling of Trotter time step with molecule size} \label{sct:app-scaling}

A key component of the overall scaling of the QFCI algorithm comes from the Trotter time step. To keep a fixed accuracy, the Trotter time step will generally have to decrease as the molecule, and hence the number of terms, grows. In this appendix, we will describe our method used to estimate this scaling.

For a given molecule, we estimate the Trotter error by finding the eigenvalues $\lambda_i$ of the unitary matrix $U$ generated by the time evolution with a given Trotter decomposition for a Trotter timestep $\Delta_t$. We can obtain an estimate for the energies from $\log(\lambda_i)/(-i \Delta_t)$; these are exactly the energy eigenvalues that will be measured with quantum phase estimation. To perform this diagonalization for large enough molecules, we need to resort to iterative diagonalization techniques such as the Arnoldi method. These methods generally only extract a few eigenvalues; to target the eigenvalues of $U$ corresponding to the ground state, we shift the Hamiltonian by an appropriate amount $\epsilon$ such that $E_0 + \epsilon > 0$; for small enough $\Delta_t$, the eigenvalue we are interested in is then the eigenvalue closest to 1 on the unit circle, and hence the eigenvalue with the largest real component.

In order to extract the scaling, we need to perform estimates for a number of different molecules; however, the exponential scaling of the classical algorithm with the number of spin orbitals as well as the large overhead of estimating the energy levels from $U$ instead of a direct calculation of the ground state, as would usually be done in FCI, severely restrict the number of spin orbitals we can study to roughly $N=24$. In order to have a sufficient number of generic molecules in the range $N=8$ to $N=24$, we generate random Hamiltonians that imitate the statistics of interaction terms found in real molecules. Specifically, we generate terms $h_{pp} c_p^\dagger c_p$, $h_{pq} c_p^\dagger c_q$, 
$h_{pqqp} c_p^\dagger c_q^\dagger c_q c_p$, $h_{pqqr} c_p^\dagger c_q^\dagger c_q c_r$ and $h_{pqrs} c_p^\dagger c_q^\dagger c_r c_s$ with the distribution functions for the parameters chosen as (here $u_{[a,b]}(x)$ is a uniform distrubtion of values between $a$ and $b$ evaluated at $x$):
\begin{eqnarray}
p(h_{pp}) &=& u_{[-10,0]}(h_{pp}) \\
p(h_{pq}) &=& u_{[-1,1]}(h_{pq}) \\
p(h_{pqqp}) &=& u_{[-0.5,0.5]}(h_{pqqp}) \\
p(h_{pqqr}) &=& \frac{1}{2 \cdot 0.2} e^{-|h_{pqqr}|/0.2} \\
p(h_{pqrs}) &=& \frac{1}{2 \cdot 0.1} e^{-|h_{pqrs}|/0.1}
\end{eqnarray}
We only keep terms compatible with particle-number conservation symmetry; additionally, we remove a fraction of the terms to mirror the fact that in a real molecule, terms may be forbidden by spatial symmetries. We generate molecules up to $N=24$, obtain their ground state energy for $N_e = N/2$ and $N_e = N/3$, as well as the respective energy estimates for a number of different values of $\Delta_t$, to obtain the error estimate $\epsilon(\Delta_t)$. The number of Trotter time steps required to reach accuracy $\epsilon_t$ is then given as
\begin{equation}
N_\text{Trotter} = \left( \frac{\epsilon(\Delta_t)}{\Delta_t^2 \epsilon_t} \right)^{1/2}
\end{equation}
assuming that $\epsilon(\Delta_t) \sim \Delta_t^2$, which we empirically confirm.

\section{Improved Trotter-Suzuki Error Bounds and Ground State Energy Error}
\label{sct:app-commutator}

Given a Hamiltonian $H=\sum_{i=1}^m H_i$, the second order Trotter-Suzuki approximation to $\exp(-i H \Delta_t)$ is given by
\begin{eqnarray}
U^{TS} &\equiv &\exp(-i H_1 \Delta_t/2) ... \exp(-i H_m \Delta_t/2) \\ \nonumber && \times \exp(-i H_m \Delta_t/2)  ...  \exp(-i H_1 \Delta_t/2).
\end{eqnarray}
The
usual derivation of the bound on second order Trotter-Suzuki error proceeds by expanding both the exact expression $U=\exp(-i H \Delta_t)$ and the second order Trotter-Suzuki approximation $U^{TS}$ to third order in $\Delta_t$ and showing that they agree at first and second order and bounding the error at third order.  The error is shown to be bounded by $(m \Lambda \Delta_t)^3$, where $\Lambda$ is an upper bound to $\Vert H_i \Vert$.  Thus, for evolution for a fixed time $t$, with time step $\Delta_t$, the error that accumulates is at most $t (m \Lambda)^3 \Delta_t^2$, meaning that to have small error at fixed $t,\Lambda$ it suffices to have $\Delta_t \ll m^{-3/2}$.  The Trotter number scales as $1/\Delta_t$ so one needs a Trotter number scaling as $m^{3/2}$.

Before giving an improved bound, let us consider the impact of this Trotter-Suzuki error on the ground state energy.  The bounds on Trotter-Suzuki error are typically expressed by stating an error estimate of the form $\Vert U - U^{TS} \Vert \leq \epsilon$, for some number $\epsilon$.  If this is the error for a single time-step $\Delta_t$, then the error for evolution over a large time $t$ may be much larger: $\Vert U^{t/\Delta_t} - (U^{TS})^{t/\Delta_t} \Vert \leq (t/\Delta_t) \epsilon$.  Hence, it might seem that since phase estimation requires evolution for $t>>\Delta_t$, we will encounter a large error.  In fact, this is not true.  Phase estimation does require evolution for large time $t>>\Delta_t$, but this is done simply to estimate eigenvalues of $U^{TS}$.  Hence, the error in energy can be directly obtained from the difference between eigenvalues of $U$ and $U^{TS}$.  We will assume that the time step is short enough that all eigenvalues of both $U$ and $U^{TS}$ lie sufficiently close to $1$ that there is no ambiguity in which branch cut of the logarithm should be used to determine the energy.
 So, the error in ground state energy is bounded by
$$\frac{1}{\Delta_t} \Vert U-U^{TS} \Vert,$$ and so for second order Trotter-Suzuki we will obtain an error in ground state energy that scales as $(m \Lambda^3) \Delta_t^2$ at worst.

However, in a quantum chemistry setting, we have the case that many of the commutators $[H_i,H_j]$ are equal to zero.   This significantly reduces the error.  Let $K$ be the maximum over $i$ of the number of terms $H_j$ which have a nonvanishing commutator with $H_i$.
For the Hamiltonian (\ref{eqn:H}), $K=\ord{N^3}$, since any two terms that do not commute must agree on at least one index. 
As a result, we can bound the error by 
\be
\label{ordbound}
\Vert U - U^{TS} \Vert \leq
\ord{m K^2 \Lambda^3 \Delta_t^3},
\ee 
rather than the previous bound of $\ord{m^3 \Lambda^3 \Delta_t^3}$,
as we now show.  
For our problem, with $m=\ord{N^4},K=\ord{N^3}$, this means that a Trotter number of $\ord{N^5}$ suffices to obtain small error.

We begin with a slightly naive derivation of the bound above using series expansions.  This bound suffers from some problems as we point out at the end, and afterwards we give a corrected proof of the result.
The series expansion up to third order can be written as a sum of many terms.  All terms cancel at first and second order.  There are terms which are third order in $H_i$ for some given $i$; these also cancel exactly.  There are terms which are first order in $H_i$ and second order in $H_j$ for some $j \neq i$.  There are at most $m k$ such terms, so they contribute at most $\ord{m k \Delta^3 \Delta_t^3}$ to the total error.  Finally, there are terms which are first order in $H_i,H_j,H_k$ for $i \neq j \neq k \neq i$.  However these terms also cancel if any one of the three operators $H_i,H_j,H_k$ commutes with the other two operators.  Thus, there are only at most $\ord{m K^2 \Delta_t^3}$ such terms.  Thus, the third order error is bounded by $\ord{m K^2 \Delta_t^3}$.

A similar calculation can be done at higher order.  
In general, it is useful to introduce a ``linked cluster expansion" to keep track of combinatorics.    Consider a term at $q$-th order which is linear in $H_{i_1},H_{i_2},...,H_{i_m}$, where the $i_a$ need not be distinct from each other.  We introduce a diagrammatic notation, writing $q$ distinct points corresponding to the terms and drawing a line between the $a$-th points and the $b$-th point if $[H_{i_a},H_{i_b}] \neq 0$.  Define the ``linked clusters" to be the connected components of the resulting graph.  A term at $m$-th order cancels unless at least one of the clusters contains at least three points.  Summing over all clusters satisfying this condition, this bounds the expression at $q$-th order by
$\ord{m^{q-2} K^2 (\Lambda \Delta_t)^q}$.

The trouble with this series expansion method is twofold.  First, there is a combinatoric issue of bounding the prefactors in front of the higher order terms. While this can be dealt with, the more serious issue is that the expansion parameter in the series expansion is actually still $m \Lambda \Delta_t$, so the bound $\ord{m K^2 \Lambda^3 \Delta_t^3}$ can only be proven for sufficiently small $m \Lambda \Delta_t$.

We now give an alternate derivation that does not assume $m \Lambda \Delta_t$ is small also.  We first bound the error in the second order Suzuki expansion for a problem with only two terms, called $A$ and $B$.  Let us fix $\Delta_t=1$.  Define $H(x)=B+(1-x)A$. 
We wish to bound
\begin{eqnarray}
\nonumber
&&\Vert \exp(-i \frac{A}{2}) \exp(-i B) \exp(-i \frac{A}{2})-\exp(-i H(0)) \Vert
\\ \nonumber &=& \Vert \int_0^1 \partial_x \Bigl(\exp(-i x \frac{A}{2}) \exp(-i H(x))
\exp(-i x \frac{A}{2}) \Bigr)  {\rm d} x \Vert.
\end{eqnarray}
We will bound the norm of the derivative on the right-hand side of the above expression, and then integrate this bound over $0 \leq x \leq 1$ to bound the
second order error.
 Let $A(t,x)=\exp(i H(x) t) A \exp(-i H(x) t)$.  Then,
\begin{eqnarray}\nonumber
&&\Vert \partial_x \Bigl( \exp(-i x\frac{A}{2}) \exp(-i H(x)) \exp(-i x \frac{A}{2}) \Bigr) \Vert \\ \nonumber
&= & \Vert \exp(-i x \frac{A}{2})  \exp(-i H(x)) \Bigl(-i\frac{A}{2}-i\frac{A(1,x)}{2} \\ \nonumber &&+i\int_0^1 A(t,x) {\rm d}t \Bigr) \exp(-i x \frac{A}{2}) \Vert \\ \nonumber
&=& \Vert -\frac{A(0,x)}{2}-\frac{A(1,x)}{2} +\int_0^1 A(t,x) {\rm d}t \Vert.
\end{eqnarray}
We have $A(t,x)=A(0,x)+t A'(0,x) + \int_0^t (t-s) A''(s,x) {\rm d}s$, where $A'$ and $A''$ represent first and second derivatives with respect to $t$ in $A(t,x)$.  One may verify that the terms in $A$ and $A'$ cancel in the above equation, leaving only terms in $A''$, giving after some calculus
\begin{eqnarray}\nonumber
&& \Vert \partial_x \Bigl( \exp(-i x \frac{A}{2}) \exp(-i H(x)) \exp(-i x \frac{A}{2}) \Bigr)\Vert  
\\ \nonumber
& = &  \Vert \int_0^1 \frac{s-s^2}{2} A''(s,x) {\rm d} s \Vert \\ \nonumber
& \leq &  \int_0^1 \Vert A''(s,x) \Vert {\rm d} s.
\end{eqnarray}
 Integrating over $x$, and using $\Vert A''(s,x) \Vert = \Vert [[A,H(x)],H(x)] \Vert \leq \Vert [[A,B],A]+[[A,B],B] \Vert$, we find that
\begin{eqnarray}
\label{diffbnd}
\nonumber
&&\Vert \exp(-i \frac{A}{2}) \exp(-i B) \exp(-i \frac{A}{2})-\exp(-i (A+B)) \Vert \\ &\leq &\Vert [[A,B],A] \Vert + \Vert [[A,B],B] \Vert.
\end{eqnarray}

We now apply Eq.~(\ref{diffbnd}) inductively to give the desired bound on $U-U^{TS}$.  Let $U_j=\exp(-i \sum_{j \leq k \leq m} H_k)$ and let 
\begin{eqnarray}
\nonumber
U_j^{TS}&=& \exp(-i H_j \frac{\Delta_t}{2}) ... \exp(-i H_m \frac{\Delta_t}{2}) \\  && \times \exp(-i H_m \frac{\Delta_t}{2})  ... \exp(-i H_j \frac{\Delta_t}{2}).
\end{eqnarray}
Then by a triangle inequality,
\begin{eqnarray}
&&\Vert U_{j-1}-U_{j-1}^{TS} \Vert \\ \nonumber &\leq & \Vert U_{j-1} -  \exp(i H_{j-1} \frac{\Delta_t}{2}) U_j \exp(-i H_{j-1} \frac{\Delta_t}{2}) \Vert \\ \nonumber && + \Vert  \exp(i H_{j-1} \frac{\Delta_t}{2}) U_j \exp(-i H_{j-1} \frac{\Delta_t}{2}) -U_{j-1}^{TS} \Vert
\\ \nonumber &= & \Vert U_{j-1} -  \exp(i H_{j-1} \frac{\Delta_t}{2}) U_j \exp(-i H_{j-1} \frac{\Delta_t}{2}) \Vert \\ \nonumber && + \Vert U_{j}-U_{j}^{TS} \Vert.
\end{eqnarray}
Using Eq.~(\ref{diffbnd}) to bound the first term on the right-hand side of the above equation, taking $A=H_{j-1} \Delta_t$ and $B=\sum_{j \leq k \leq m} H_k \Delta_t$, we get
\be
\Vert U_{j-1}-U_{j-1}^{TS} \Vert \leq {\rm const.} \times \Lambda^3 \Delta_t^3 K^2 + \Vert U_{j}-U_{j}^{TS} \Vert,
\ee
where the constant is a numeric constant of order unity.
Summing over $j$, we obtain the desired bound Eq.~(\ref{ordbound}) for $\Vert U-U^{TS} \Vert = \Vert U_1 - U_1^{TS} \Vert$.

It seems likely that this same proof will also work to bound higher order errors in higher order Trotter-Suzuki expressions.  For $2r$-th order Trotter-Suzuki, we expect to bound the error by $\ord{m k^{2r} \Lambda^{2r+1} \Delta_t^{2r+1}}$.  If indeed this holds, for $K=N^3$ and $m=N^4$, for sixth order Trotter-Suzuki the Trotter number required to obtain given error is actually {\it sublinear} in $m$.
\section{Coalescing}
\label{sec:coalescing}
Our work makes it clear that algorithmic improvements are needed to make quantum chemistry practical on a quantum computer.  
Here we suggest one potential improvement which signficantly decreases the total number of gate operations required by using an adaptive Trotter decomposition for terms of different magnitude.  Unfortunately,
there is some tradeoff in this scheme as the rearrangment we suggest likely increases the Trotter error.  Further research is required to understand, in what regimes, this tradeoff is such that this is a beneficial approach.

The key idea is based on the observation that in the Hartree-Fock basis,
many off-diagonal matrix elements are very small. In the unitary time
evolution these small terms can be applied with much larger Trotter time 
steps $\Delta_t$ compared to the larger terms. As a consequence fewer
circuits have to be applied and the total gate count is significantly reduced.

Specifically, we propose to use a different time step $\Delta_t^{(k)}$ for each of
the terms in the Hamiltonian $h_k$ and to choose them such that the product
of the amplitude of the term and the time step $\Delta_t^{(k)} \cdot \parallel h_k \parallel$
is roughly homogeneous across all terms.

Equivalently, consider the Trotter-decomposed time evolution operator for a
given total time $T$,
\begin{equation}
U = \left( \prod U_k \right)^{T/\Delta_t},
\end{equation}
where $U_k$ applies the Hamiltonian term $h_k$ for a Trotter step $\Delta_t$.
In the Hartree-Fock basis, many of these terms are extremely close to the identity.
We can then imagine rearranging the terms in this expansion such that identical
terms that are very close to the identity, i.e. that have a very small coefficient in
the Hamiltonian, are grouped together and can be executed at once.

From this description it becomes clear that some additional discretization error
will be accumulated by changing the order of terms, and it may therefore be
necessary to reduce the Trotter time step to keep the total error fixed. This leads
to a trade-off between grouping terms and keeping the error constant. The best
scheme within this trade-off depends sensitively on the specific distribution of
Hamiltonian terms $h_k$. In this paper, we do not address this question in detail,
but defer it to future work.  However, we do give a brief theoretical analysis of one simple coalescing scheme.

Let the Hamiltonian $H$ be a sum of terms $H=\sum_i H_i$.  Divide these terms $H_i$ into {\it buckets}, so that every term is in exactly one bucket.
Label the buckets $1,...,k$, and let $B_a$ be the set of integers $i$ so that $H_i$ is in bucket $a$.
Assume that there are $N_a$ terms in bucket $a$ with $\Vert H_i \Vert \leq \Lambda_a$ for $i \in B_a$.  We arrange the buckets to contain terms in decreasing order of magnitude, so that $B_1$ contains the largest terms and is executed the most frequently, while later buckets are exectued less frequently.

We wish to approximate
$U=\exp(-i H t)$
with a quantum circuit, and we assume that we have circuits to implement $\exp(-i \theta H_i)$ for any $i$.

We analyze a coalescing scheme to approximat this unitary, and show that it achieves error $\epsilon$ in operator norm bounded by
\be
\label{coalmain}
\epsilon \leq 
\sum_{a}  O(\frac{ S_{a}^3+T_{a-1} S_a^2 + T_{a-1}^2 S_a }{(2^{k-a})^2}),
\ee
where we define
\be
S_a=N_a \Lambda_a t
\ee
and
\be
T_a = \sum_{b\leq a} S_b,
\ee
with $T_0=0$.
The scheme is defined by Eqs.~(\ref{Wadef},\ref{V1def},\ref{VInddef}), with $V_k$ defined by those equations being the approximation to $U$.

To gain some intuition for Eq.~(\ref{coalmain}), let $K_a=\sum_{i \in B_a} H_i$ be the sum of all terms in bucket $a$.
Then, the first term $O(S_a^3/(2^{k-a})^2$) in the equation is the error we obtain by approximating
$\exp(-i K_a t)$ by
doing a
a second order Trotter-Suzuki expansion to $\exp(-i K_a \frac{t}{2^{k-a}})$ and then taking the $2^{k-a}$-th power of that approximation.  In a sense, this first term results from errors in commutators in a single bucket.  The terms in $T_{a-1} S_a^2$ and $T_{a-1}^2 S_a$ result from interaction between buckets.  The terms resulting from interaction between buckets have one unfortunate effect: if there is a large term in the first bucket (which gives a large $T_a$ and which will be approximated with a very short Trotter step), then ideally we would like to suppress the error by a large denominator $(2^{k-1})^2$, and indeed the first term in Eq.~(\ref{coalmain}) is suppressed by this factor.  However, since this term interacts with terms in later buckets, there will be error terms with denominators of order $(2^{k-a})^2$ for all $a$ also appearing in the second and third terms in Eq.~(\ref{coalmain}) and this is a smaller denominator.  However, these terms will be at most second order in $T_1$ and will be at least first order in $S_a$ so if $S_a$ is small then this will help suppress these terms.

To define and analyze the scheme, let
\be
J_a=\sum_{b \leq a} \sum_{i \in B_b} H_i
\ee
be the sum of all terms in buckets $1,...,a$ and
let
\be
U_a=\exp(-i J_a \frac{t}{2^{k-a}}),
\ee
so that $U=U_k$.
We proceed inductively, using an approximation to $U_a$ to construct an approximation to $U_{a+1}$.
We call this approximation $V_a$.
and let
\be
\epsilon_a = \Vert V_a - U_a \Vert.
\ee
We define $V_1$ by a standard second order Trotter-Suzuki as follows.  Let
\be
\label{Wadef}
W_a=\prod_{i \in B_1} \exp(-i \frac{1}{2} H_i \frac{t}{2^{k-a}}),
\ee
where the product is taken in any fixed arbitrary order, and let $W'_a$ denote the same product as $W_a$ except taken in the reverse order (if $H$ is real, then $W'_a=W_a^T$).
Then, let
\be
\label{V1def}
V_1=W_1 W'_1.
\ee
By standard estimates,
\be
\epsilon_1 \leq O(\frac{S_1}{2^{k-1}})^3.
\ee
This estimate for $\epsilon_1$ is based on a third order Taylor expansion; the lower order terms in the Taylor expansion cancel, and the higher order terms in the Taylor expansion are higher order in $S_1$.

We then define
\be
\label{VInddef}
V_{a+1} = W_{a+1} V_a^2 W'_{a+1}.
\ee
Note that
\be
\label{tri1}
\Vert V_{a+1} - W_{a+1} U_a^2 W'_{a+1} \Vert \leq 2 \epsilon_a.
\ee
Also,
\begin{eqnarray}
\label{tri2}
&& \Vert U_{a+1} - W_{a+1} U_a^2 W'_{a+1} \Vert \\ \nonumber &\leq &O(\frac{ S_{a+1}^3+\Vert J_a \Vert S_{a+1}^2 + \Vert J_a \Vert^2 S_{a+1} }{(2^{k-a-1})^3}),
\end{eqnarray}
as can be estimated using a Taylor series similar to before.
So, by Eqs.~(\ref{tri1},\ref{tri2}),
\begin{eqnarray}
\epsilon_{a+1} &\leq 2& \epsilon_a +O(\frac{ S_{a+1}^3+T_a S_{a+1}^2 + T_a^2 S_{a+1} }{(2^{k-a-1})^3}),
\end{eqnarray}
using the fact
that $\Vert J_a \Vert \leq T_a$.

Hence,
\be
\label{eabd}
\epsilon_a \leq \sum_{b \leq a} 2^{a-b} O(\frac{ S_{b}^3+T_{b-1} S_b^2 + T_{b-1}^2 S_b }{(2^{k-b-1})^3}),
\ee
and
\be
\epsilon_k \leq \sum_{a}  O(\frac{ S_{a}^3+T_{a-1} S_a^2 + T_{a-1}^2 S_a }{(2^{k-a})^2}).
\ee

\section{Circuits}
\label{sec:circuits}
Following the approach set out in Ref.~\cite{whitfield2011}, we convert each term in Equation \ref{eqn:H} via a Jordan-Wigner transformation which expresses fermionic operators in terms of Pauli spin operators \{${\sigma^x, \sigma^y, \sigma^z, \mathbf{1}}$\}. This technique allows us to preserve the necessary commutation relations by entangling all of the spin-orbitals (qubits) between annihilation and creation operators with $\mathbf{CNOT}$ gates, performing the unitary propagator ($\exp(-iHt)$) at a given strength ($t_{pq}$ or $V_{pqrs}$) and then unentangling by reversing the order of the applied $\mathbf{CNOT}$s. If the propagator is in the computational basis $\sigma^z$, then a rotation of the desired strength may be applied directly. If the propagator is in the $\sigma^x$ basis then the basis must be flipped from $\sigma^x$ to $\sigma^z$ with a Hadamard transformation $\mathbf{H}$; the operation is then performed and then returned from $\sigma^z$ to $\sigma^x$ with an additional $\mathbf{H}$ at the end. Likewise, $\mathbf{Y}$ and $\mathbf{Y^\dagger}$ may be used to flip between $\sigma^y$ and $\sigma^z$. The basis flip operations are:
\begin{eqnarray}
\mathbf{H}=\frac{1}{\sqrt{2}}\left[
    \begin{array}{cc}
    1 & 1 \\
    1 & -1
    \end{array}
    \right] \ \ \ \
    \mathbf{Y}=\frac{1}{\sqrt{2}}\left[
    \begin{array}{cc}
    1 & i \\
    i & 1
    \end{array}
    \right]
\end{eqnarray}
\vspace{0mm}

\begin{figure*}
  \includegraphics{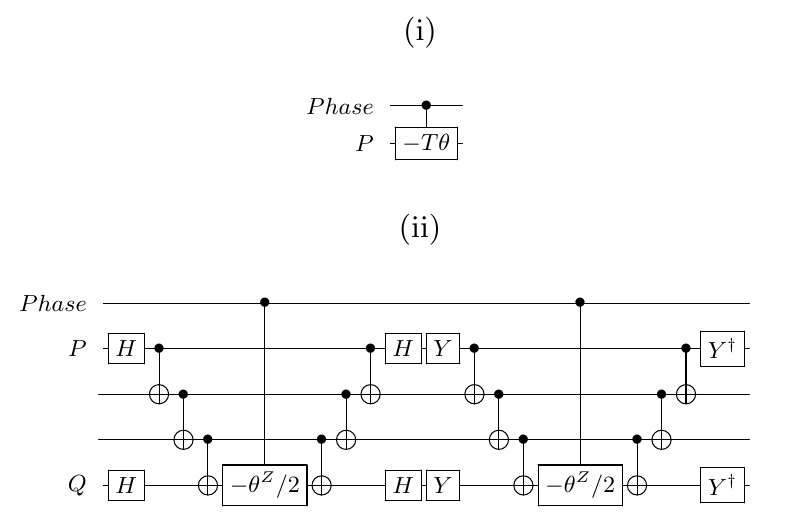}
  \caption{Circuit representations of Hamiltonian terms $H_{pp}$ and $H_{pq}$. \label{fig:circuits1} }
\end{figure*}

\begin{figure*}
  \includegraphics{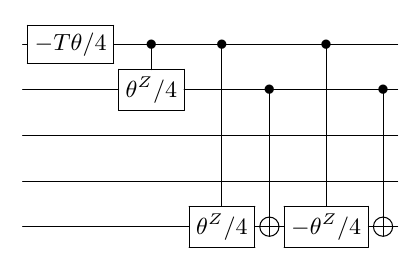}
  \caption{Circuit representation of Hamiltonian terms $H_{pqqp}$. \label{fig:circuits2} }
\end{figure*}

\begin{figure*}
  \includegraphics{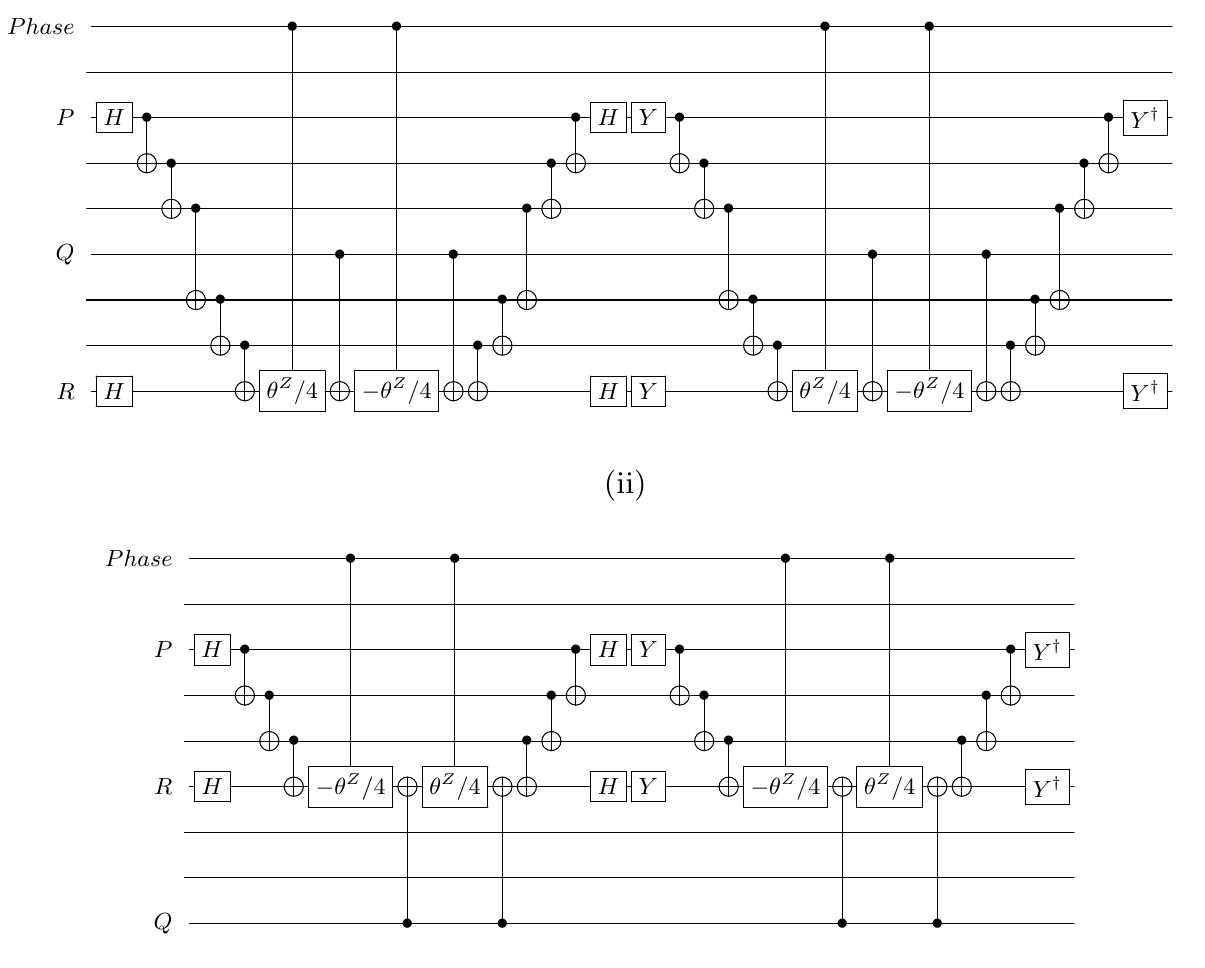}
  \caption{Circuit representations of Hamiltonian terms (i) $H_{pqqr}$, $p < q < r$, and (ii) $H_{pqqr}$, $q < p$ or $q < r$. \label{fig:circuits3} }
\end{figure*}

\begin{figure*}
  \includegraphics{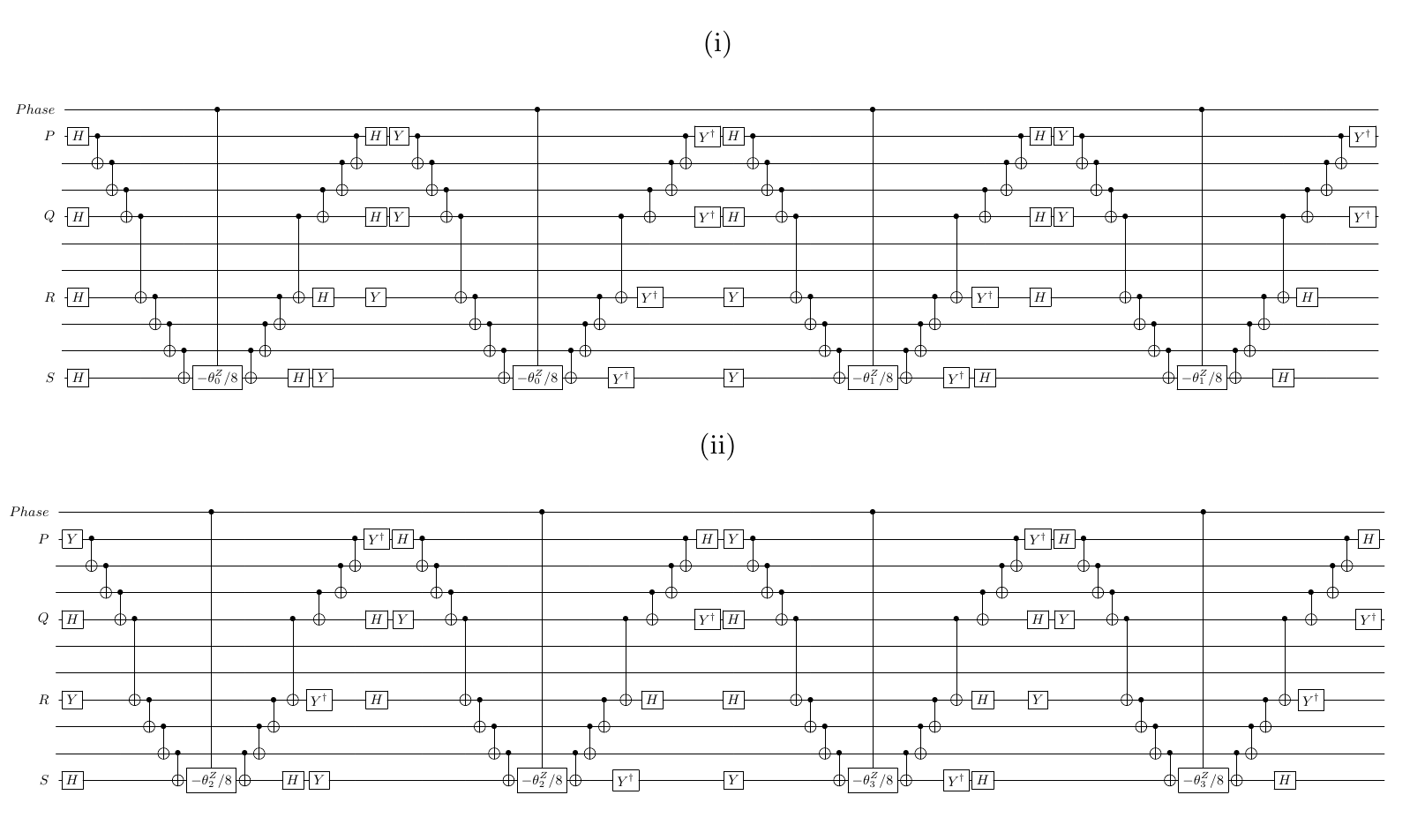}
  \caption{Circuit representation of Hamiltonian terms $H_{pqrs}$. \label{fig:circuits4} }
\end{figure*}

The resulting circuits are presented in Figs.~\ref{fig:circuits1}, \ref{fig:circuits2}, \ref{fig:circuits3} and \ref{fig:circuits4}. These circuits coincide with those of Ref.~\cite{whitfield2011}, except for those for $H_{pqqr}$ which were not needed in the example discussed in the reference. In all circuits, $\theta$ refers to the term strength ($t_{pq}$ or $V_{pqrs}$) computed by a standard technique. In our case this was PyQuante~\cite{PyQuante} running a Restricted Hartree-Fock model.

The circuits shown above are easily parameterized to show expected resource usage as much larger molecules are modeled. Table \ref{fig:circuitdepth} shows exact counts for both sequential and parallel implementation of all one and two body terms used in the the Hamiltonian.

\begin{table*}
  \begin{tabular}{|c||c|c|c|c|c||c|}
  \hline
  Sequential Circuit &Global $\mathbf{R}_z$ & $\mathbf{H}$, $\mathbf{Y}$, $\mathbf{Y}^\dag$ &$\mathbf{CNOT}$ & $\mathbf{CR}_z$  & BSM & Total \\ \hline
$H_{pp}$ & & & & 1 && 1 \\
$H_{pq}$ & & 8 & $4(q-p)$ & 2 && $10+ 4(q-p)$\\
$H_{pqqp}$ &1 & & 2 & 3 && 1+5\\
$H_{pqqr}$, $p<q<r$ & &8 & $4(r-p)$ & 4 &&  $12+4(r-p)$\\
$H_{pqqr}$, $q <p$ or $r< q$  & &8 & $4(r-p+1)$ & 4 &&  $16+4(r-p)$\\
$H_{pqrs}$ & &$8\cdot8$ & $8\cdot2(q-p+s-r+1)$ & $8\cdot 1$ && $8\cdot 9 + 8\cdot2(q-p+s-r+1)$ \\
\hline
  \hline
Parallel Circuit &Global $\mathbf{R}_z$ & $\mathbf{H}$, $\mathbf{Y}$, $\mathbf{Y}^\dag$ &$\mathbf{CNOT}$ & $\mathbf{CR}_z$  & BSM & Total \\ \hline
$H_{pp}$ & & & & 1 & & 1 \\
$H_{pq}$ & & 8 & $4$ & 2 & 4 & $18$\\
$H_{pqqp}$ &1 & & 2 & 3 & & 1+5\\
$H_{pqqr}$ & &4 & $8$ & 4 & 4 &   $24$\\
$H_{pqrs}$ & &$8\cdot2$ & $8\cdot2$ & $8\cdot 1$ & $8\cdot2$ & $8\cdot7$ \\
\hline
  \end{tabular}
  \caption{Circuit depths of the circuits for the individual terms in the Hamiltonian for sequential (top) and parallel (bottom) execution of the circuits. The global rotation gate in the $H_{pqqp}$ term needs to be counted only once for all such terms. The eight terms in $H_{pqrs}$ can be reduced to 4 or 2 for terms where some of the angles $\theta_i$ are 0 due to symmetry. For the parallel circuits, the method of Ref.~\cite{CodyJones2012} can execute the Jordan-Wigner strings in constant time at the cost of additional Bell-state measurements (BSM). }
     \label{fig:circuitdepth}
 \end{table*}

\narrowtext
\bibliographystyle{apsrev4-1}
\bibliography{paper}

\end{document}